\documentclass[a4paper,11pt]{article}
\usepackage{pos}
\PassOptionsToPackage{colorlinks,linkcolor={blue},citecolor={blue},urlcolor={red}}{hyperref}
\usepackage{hyperref}
\usepackage{natbib}
\usepackage{wrapfig}
\usepackage{float}
\usepackage{color}
\usepackage{graphicx}
\usepackage{gensymb}
\usepackage{amsmath}
\usepackage{enumitem}
\usepackage{url}
\usepackage{multirow,helvet}
\usepackage{todonotes}

\def\mnras{MNRAS}
\def\apj{ApJ}

\def\aap{A\&A}

\def\apjl{ApJL}
\def\apjs{ApJS}
\def\prd{PRD}
\def\araa{ARAA}
\def\baas{BAAS}
\def\ssr{SSR}
\def\jcap{JCAP}

\title{Pulsar Wind Nebulae (PWNe) - A Review}
\author*[a]{Jordan Eagle}

\affiliation[a]{NASA Goddard Space Flight Center,\\
Greenbelt, Maryland, USA}

\emailAdd{jordanlynneagle@gmail.com}

\abstract{Pulsar Wind Nebulae (PWNe) are relativistic, magnetic winds comprised of radiating electrons and positrons, powered by an energetic pulsar. The pulsar continuously injects particles into the PWN that are accelerated at the termination shock. As the relativistic particles enter the PWN, they radiate away the energy received at the shock as they interact with the PWN environment, generating synchrotron emission from interactions with the magnetic field of the PWN and Inverse Compton Scattering (ICS) from interactions with the local photon fields. Synchrotron emission is observed from the majority of known PWNe from radio to X-ray energies, and the ICS is observed in the $\gamma$-ray bands, from MeV to TeV energies. The particle acceleration processes at the termination shock and elsewhere within the PWN remain to be understood. Recent progress in theoretical studies have provided the capability to explain broadband observations of several PWNe including their spectral and spatial features. This work reviews some of the most compelling outcomes of recent literature, outlining the outstanding questions that remain to be answered, and how the future prospects of $\gamma$-ray astronomy will be instrumental in advancing the current understanding of PWNe.}

\FullConference{
Multifrequency Behaviour of High Energy Cosmic Sources -- XV (MULTIF2025)\\
9--14 June, 2025\\
Mondello, Palermo, Italy\\}

\begin{document}
\maketitle

%%%%%%%%%%%%%%%%%%%%%%%%%%%%%%%%%%%%%%%%%%%%%
%to-do
% consider updating some of the lower resolution images
%%%%%%%%%%%%%%%%%%%%%%%%%%%%%%%%%%%%%%%%%%%%%

\section{Introduction}
\begin{figure*}
\begin{minipage}[b]{.5\textwidth}
\centering
\includegraphics[width=0.75\linewidth]{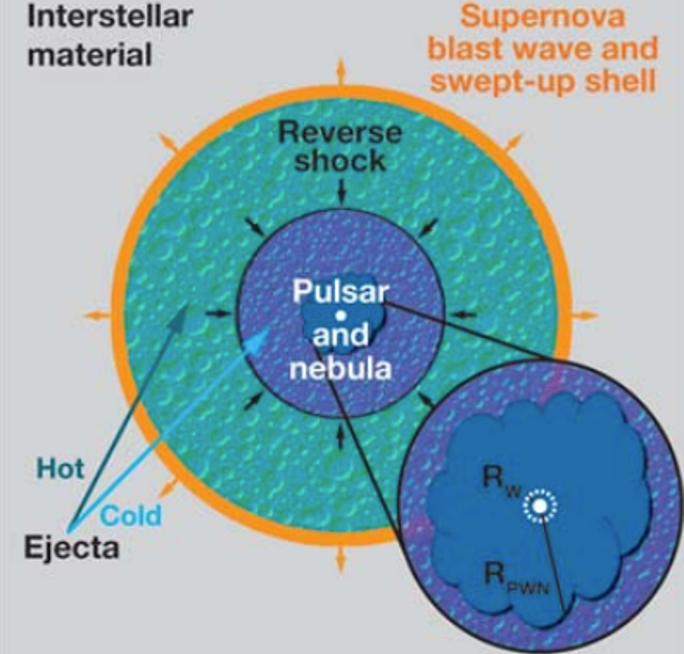}
\end{minipage}
\begin{minipage}[b]{.5\textwidth}
\hspace{-1cm}
\includegraphics[width=1.1\linewidth]{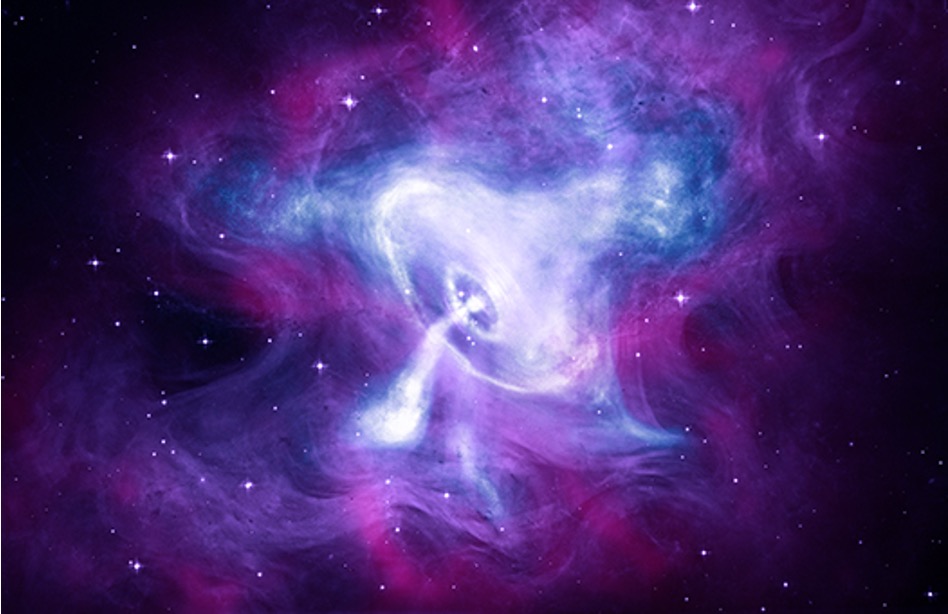}
\end{minipage}
\caption{{\it Left:} An illustration of the primary layers that make up a core collapse SNR from \citep{gaensler2006}. The SN blast wave and swept up shell is the outer shell that expands into the interstellar medium (ISM). At later times, a reverse shock is generated from the outer shell that accelerates towards the center of the SNR where the central pulsar and PWN are located. The reverse shock shocks and heats the colder SN ejecta in the SNR interior, forming the separate layers of the ejecta ``hot`` (shocked by the reverse shock) and ``cold`` (unshocked). At early times, the PWN expands into the cold SNR interior. At later times, it interacts with both shocked SN ejecta and the reverse shock. {\it Right:} The Crab Nebula in X-ray (blue, NASA/CXC/SAO), optical (red, NASA/STScI), and infrared (pink, NASA-JPL-Caltech). The pulsar is observed as the central point source encompassed by two rings. The innermost ring is corresponds to the predicted location of the termination shock. The region beyond it comprises the radiating PWN, featuring unique structures such as the larger toroidal ring, filaments,  and a jet emanating from the poles of the pulsar.}\label{fig:pwn_intro}
\vspace{-0.5cm}
\end{figure*}

%Pulsar Wind Nebulae (PWNe) are energetic, magnetic, and relativistic winds comprised of radiating electrons and positrons that are continuously supplied by an energetic, rapidly rotating neutron star. 
Descending from core collapse supernovae (SNe), pulsar wind nebulae (PWNe) are found located at or near the SN explosion center, encompassing the central pulsar, and are altogether bound by the supernova remnant (SNR) interior and exterior shell, see Figure~\ref{fig:pwn_intro}, left panel. The central pulsar continuously injects relativistic electrons and positrons into the PWN that are accelerated at the termination shock. As the relativistic particles enter the PWN, they radiate away the energy received at the shock as they interact with the PWN environment, generating synchrotron emission from interactions with the magnetic field of the PWN and Inverse Compton Scattering (ICS) from interactions with the local photon fields. Synchrotron emission is observed from the majority of known PWNe from radio to X-ray energies, and the ICS is observed in the $\gamma$-ray bands, from MeV to TeV energies. As such, PWNe are excellent candidates for Galactic cosmic ray (CR) acceleration and may be responsible for local enhancements observed in the electron-positron flux \citep{crflux}. 

The most famous and best-studied examples of PWNe is the Crab Nebula, shown in the right panel of Figure~\ref{fig:pwn_intro}. The innermost ring of the Crab Nebula in the right panel of Figure~\ref{fig:pwn_intro} corresponds to the theoretical expectation of the location of the termination shock. The site of the termination shock may facilitate efficient particle acceleration through diffusive shock acceleration (DSA). The conditions that enable DSA as well as the characteristics of other acceleration mechanisms and regions within the PWN are poorly understood. Recent progress in theoretical studies of these systems have provided the capability to explain broadband observations of several PWNe including their spectral and spatial features, incorporating multiple acceleration processes and sites. 

\begin{figure*}
\begin{minipage}[b]{1.0\textwidth}
\centering
\includegraphics[width=0.75\linewidth]{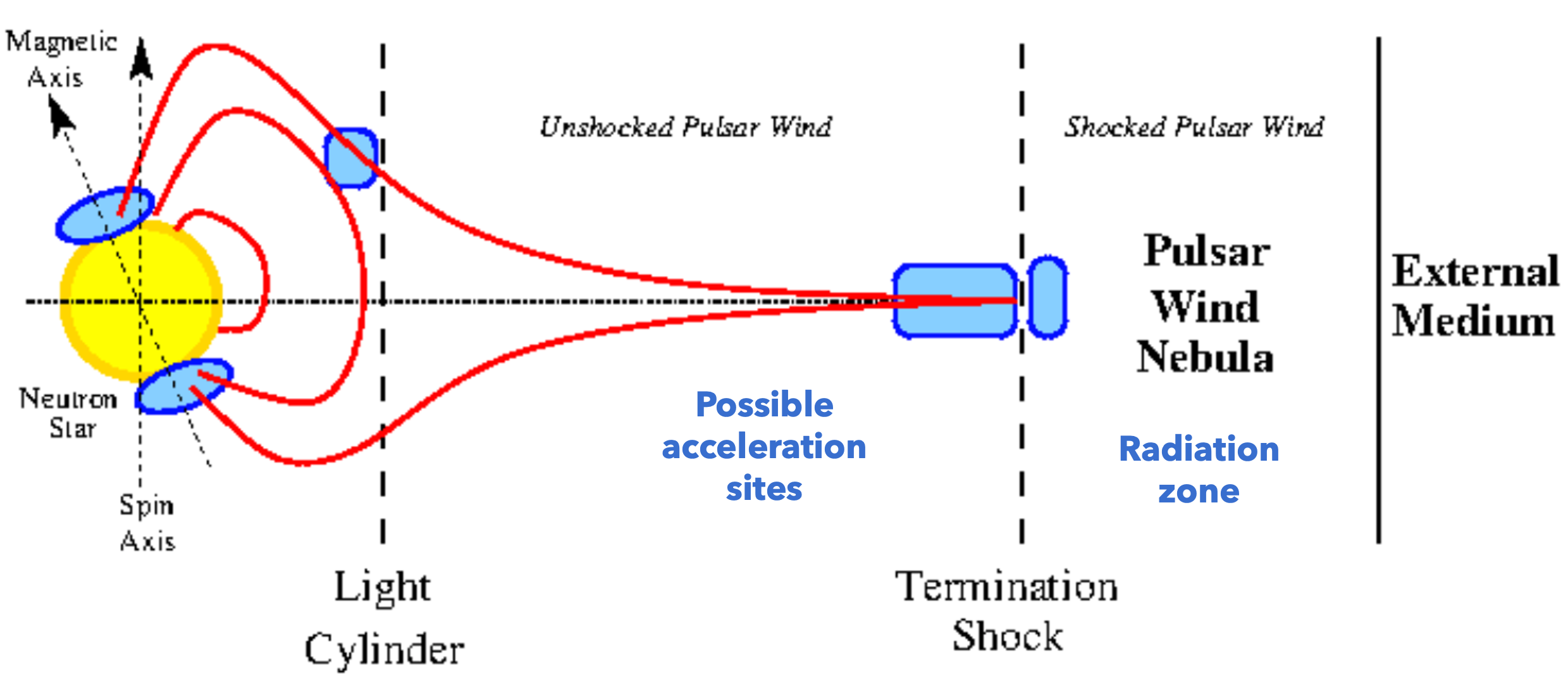}
\end{minipage}
\caption{A diagram from \citep{gelfand2019} demonstrating the basic features of a pulsar-PWN system, highlighting the possible sites of efficient particle acceleration in blue.}\label{fig:acceleration_intro}
\vspace{-0.5cm}
\end{figure*}

In this work, we focus on the possible acceleration sites that can reproduce the observations of PWNe and are highlighted in Figure~\ref{fig:acceleration_intro}. In Section~\ref{sec:cr_acc}, we summarize the most promising sites of particle acceleration around pulsars and within PWNe. In Section~\ref{sec:rad_prop}, we describe the radiative properties of the Crab Nebula and the recent theoretical works that can reproduce them using the frameworks introduced in Section~\ref{sec:cr_acc}. A discussion on the effects of evolution on the radiative properties of PWNe is also presented with a focus on the expected $\gamma$-ray signatures from evolved PWNe. In Section~\ref{sec:gamma_searches}, we provide an overview of several case studies involving $\gamma$-ray bright PWNe. In Section~\ref{sec:gamma_future}, we outline the next generation of $\gamma$-ray astronomy observatories and their potential to reveal remaining mysteries around the nature of PWNe and provide our conclusions in Section~\ref{sec:conclusion}.

\section{Cosmic Ray Acceleration}\label{sec:cr_acc}
\begin{figure*}
\begin{minipage}[b]{.3\textwidth}
\hspace{-1cm}
\includegraphics[width=1.1\linewidth]{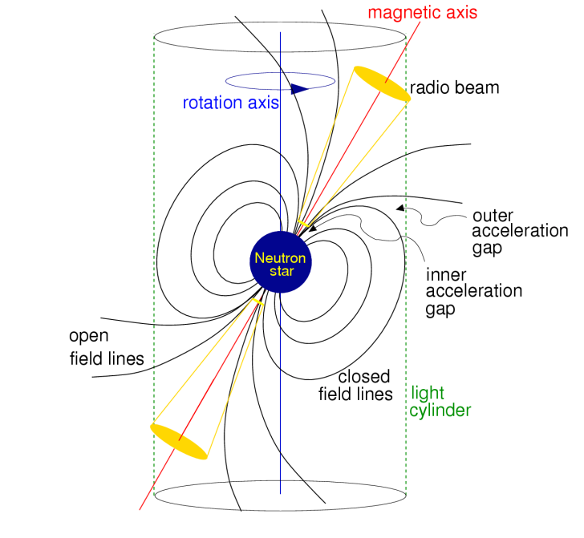}
\end{minipage}
\begin{minipage}[b]{.3\textwidth}
\hspace{-0.65cm}
\includegraphics[width=1.1\linewidth]{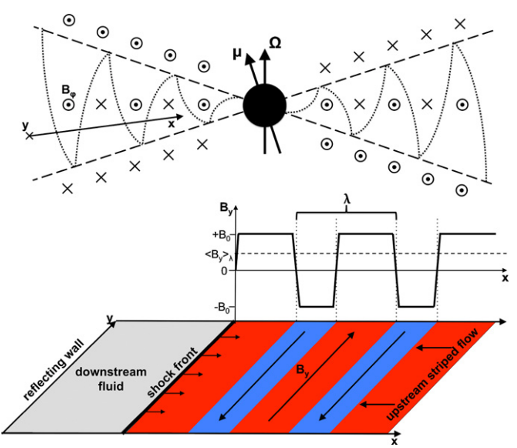}
\end{minipage}
\begin{minipage}[b]{.3\textwidth}
\hspace{-0.50cm}
\includegraphics[width=1.2\linewidth]{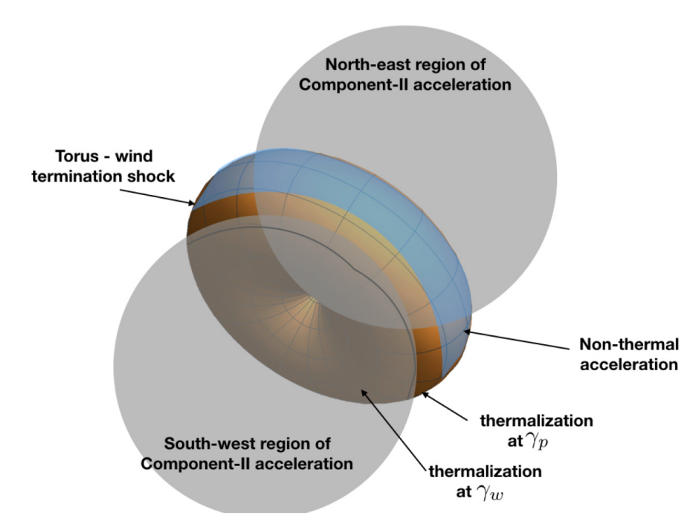}
\end{minipage}
\caption{{\it Left:}  A digram of the magnetosphere and possible acceleration regions from \citep{lorimer2012}. {\it Middle:} A proposed structure of the equatorial region of the unshocked pulsar wind  (the striped wind model) \citep{sironi2011}. {\it Right:} The presented structure of the Crab nebula involving acceleration processes in both the termination shock in the equatorial region and in the polar regions of shocked pulsar wind \cite{lyutikov2019}.}\label{fig:cr_acc_intro}
\vspace{-0.5cm}
\end{figure*}

Efficient particle acceleration, producing cosmic rays (CRs) up to PeV energy, is expected to occur within energetic environments in the Milky Way Galaxy such as SNRs, pulsars, and PWNe. Cosmic ray acceleration produced by pulsars is most likely occurring within the pulsar magnetosphere, which is located within the light cylinder, a region of space near the pulsar and defined by the co-rotational velocity of the magnetic field and particles with respect to the neutron star surface being equal to the speed of light \citep{lorimer2012}. There are gaps between the magnetic field lines near the neutron star surface (``inner gap'') and near the light cylinder radius (``outer gap'') that may facilitate efficient particle acceleration, see Figure~\ref{fig:cr_acc_intro}, left panel. Theoretical models of the pulsar magnetosphere indicate that the outer acceleration gap generates pulsed $\gamma$-ray emission from the relativistic electrons and positrons accelerating in the open magnetic field line region. On the other hand, radio emission may be generated near the polar regions where the inner acceleration gap is located, closer to the neutron star surface and among the closed magnetic field lines. 

As the relativistic particles exit the magnetosphere, it becomes a highly magnetized plasma, with the majority of the energy being carried in the magnetic field of the plasma.  Many pulsars have a rotational axis that is misaligned to their magnetic field axis, which leads to the highly magnetized plasma forming equatorial current sheets between alternating magnetic field polarity, the so-called ``striped wind model``, see Figure~\ref{fig:cr_acc_intro}, middle panel. The alternating current sheets exist between the pulsar magnetosphere and the termination shock of the PWN \citep{mitchell2022}. The termination shock is the natural consequence of the relativistic plasma expanding into a nonrelativistic medium. Many theoretical models and observations of PWNe suggest that at the termination shock, the relativistic plasma transitions from being dominated in magnetic energy to dominated in particle energy and the striped wind model offers a plausible environment for this to occur \citep{sironi2011}. In the striped wind model, magnetic energy is dissipated by shock compression of the alternating magnetic field polarity and current sheets, causing magnetic reconnection and enabling energy transfer to the particles. Moreover, the striped wind model provides a plausible solution to the sigma problem, where sigma is defined as the ratio between the Poynting flux and the particle flux of the relativistic plasma. At the magnetosphere, $\sigma >> 1$, but PWN observations as well as the current framework of DSA require $\sigma << 1$ at the termination shock to facilitate efficient particle acceleration \citep[e.g.,][]{mitchell2022}.

\begin{wrapfigure}{r}{0.6\textwidth}
\begin{minipage}{1.0\linewidth}
\hspace{-0.55cm}
\includegraphics[width=1.0\linewidth]{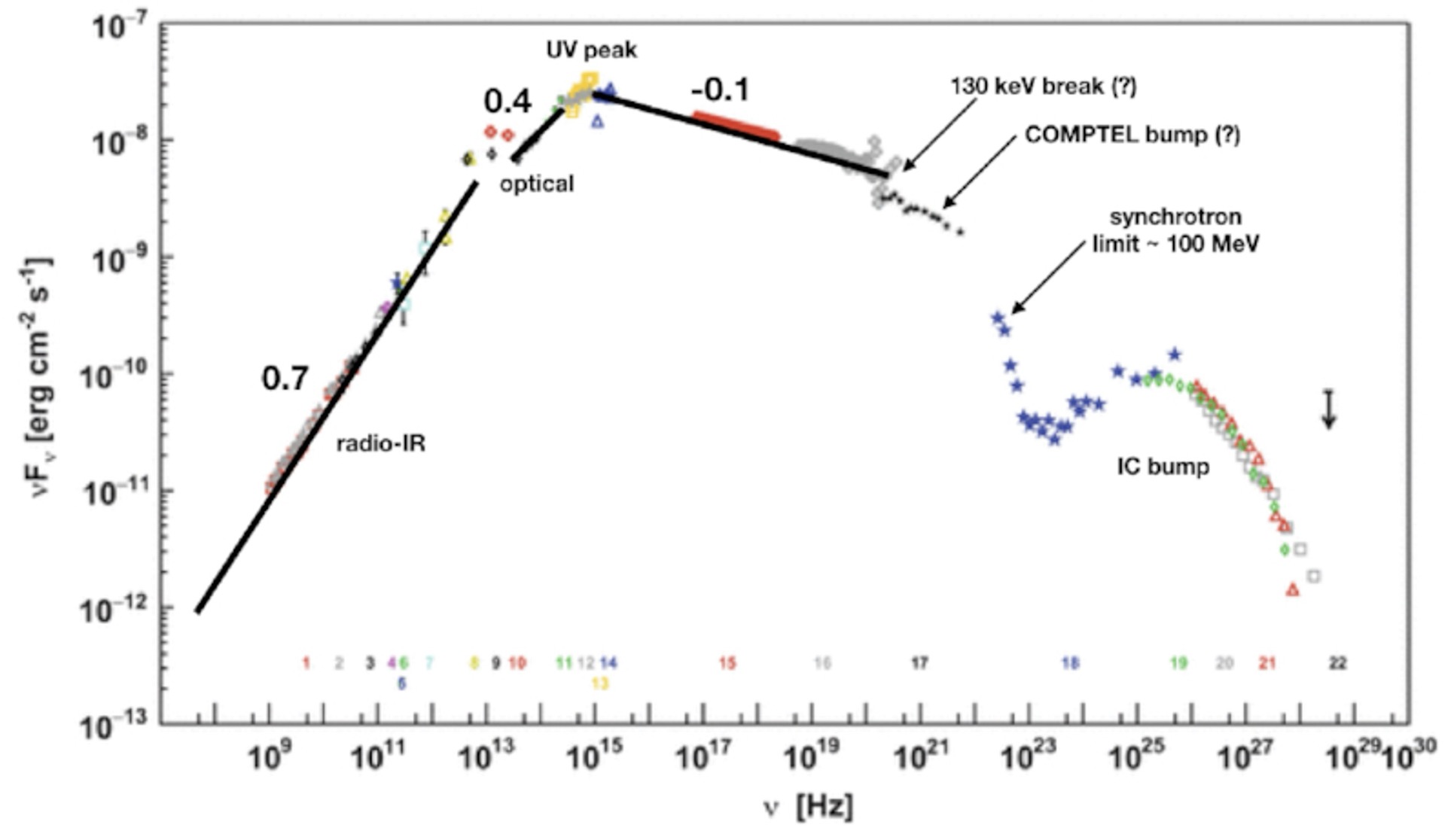}
\end{minipage}
\caption{The broadband data and spectral model of the Crab Nebula from \citep{lyutikov2019}.}\label{fig:crab_spec}
\vspace{-.50cm}
\end{wrapfigure}

Particle-in-cell (PIC) simulations suggest that DSA (also called the Fermi process) may not be the only acceleration mechanism occurring at the shock boundary. The findings also indicate that efficient particle acceleration is not confined to only the termination shock of a PWN \citep[e.g.,][]{sironi2008,sironi2017}. In general, DSA theory cannot easily explain observations of PWNe alone. As an example, the broadband spectrum of the Crab nebula and the implied electron distribution deviates from the DSA framework. A single power-law particle injection spectrum with an index greater than 2 is expected if particles are undergoing DSA, however, the observed spectrum from the Crab implies a particle index of $\sim 1.5$. 
Recent work \citep{lyutikov2019} has shown that two electron populations can explain the broadband spectrum, one which could be undergoing efficient particle acceleration via the Fermi process in the equatorial regions (i.e., at the termination shock) and a second component where efficient particle acceleration occurs from magnetic reconnection layers within the nebula, see the right panel of Figure~\ref{fig:cr_acc_intro}. 
In this interpretation, the synchrotron radiation can be explained from particles undergoing Fermi acceleration and subsequently dominating the 0.1--10\,keV X-ray range. Thereafter the magnetic reconnection layers contribute to particle acceleration, producing the MeV bump first observed by the Compton Telescope (COMPTEL). The broadband spectrum of the Crab is shown in Figure~\ref{fig:crab_spec}, adapted from \citep{lyutikov2019}. A summary of the particle characteristics invoked in \citep{lyutikov2019} is reported in Table~\ref{tab:crab}.

\begin{table}
\centering
\begin{tabular}{| c || c c |}
\hline
\ Property & Population 1 & Population 2 \\
\hline 
\hline 
\  Particle Injection Index & 2.2 & 1.6 \\
\hline 
\ Physical region & Equatorial & Polar \\
\hline 
\ Acceleration mechanism & DSA & Turbulence-driven \\
\ Spectral Features & X-ray and $\gamma$-ray & Steep radio spectrum and $\gamma$-ray flares \\
\hline 
\end{tabular}
\caption{Characteristics of the two electron populations invoked in \citep{lyutikov2019} that explain the broadband observations of the Crab Nebula.}
\label{tab:crab}
\end{table}

In the polar regions of the shocked pulsar wind, it is possible for turbulence-driven acceleration to efficiently produce CRs. Similar to the turbulence generated by the driven reconnecting magnetic fields at the termination shock, magnetized, turbulent hot spots can form within the PWN that drive efficient particle acceleration. Here, magnetized turbulence can also generate reconnecting magnetic fields and may be produced, at least in part, by instabilities such as the Weibel instability \citep[e.g.,][]{sironi2011}. The Weibel instability is essentially the magnetized turbulence generated as a natural consequence to the relativistic motion of charged particles and may be sufficient to drive particle acceleration. 

\begin{wrapfigure}{l}{0.5\textwidth}
\begin{minipage}{1.0\linewidth}
\vspace{-0.65cm}
\hspace{-0.35cm}
\includegraphics[width=1.0\linewidth]{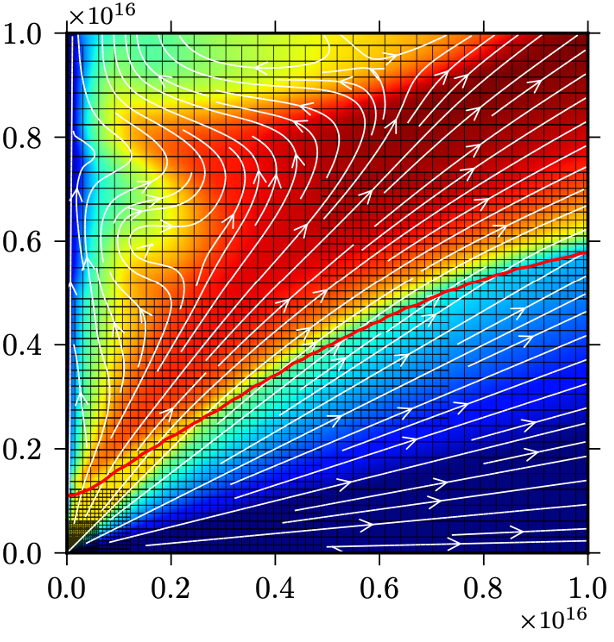}
\end{minipage}
\vspace{-.250cm}
\caption{A slice of the physical region in cm near the termination shock (red line) shown as a colormap in $\log_{10}{\sigma} = [-2,0.8]$ (from blue to red for lower to higher values) from a simulation reported in \citep{porth2014}.}\label{fig:porth_2014}
\vspace{-0.5cm}
\end{wrapfigure}

In the polar regions of the PWN, it is not required to have $\sigma << 1$ throughout. Some hot spots may actually require high magnetization with $\sigma >> 1$, leading to steeper particle spectra (particle injection index approaching 1). Averaging the high $\sigma$ regions with other weakly magnetized regions leads to an overall softer $p$ \citep{porth2014}. As a result, turbulence-driven acceleration in the polar regions of the PWN can explain the steep radio spectra typically observed by PWN including the Crab (Table~\ref{tab:crab}), see Figure~\ref{fig:porth_2014}. 

\section{Radiative Properties}\label{sec:rad_prop}

In order to accurately determine the acceleration mechanisms and regions of a PWN, the influences of PWN age on the radiative properties must be considered. The Crab Nebula is a relatively young system, $\tau \sim 1$\,kyr, but other PWNe that are much older, such as Vela-X with an age nearly ten times as old as the Crab, $\tau \sim 10$\,kyr, has similar spectral and spatial features, see Figure~\ref{fig:vela}.

Observations of evolved PWNe like Vela-X suggest multiple electron populations forming multiple emission regions \citep[e.g.,][]{dejager2008, hinton2011}. The cocoon of Vela-X represents a compact nebula observed in X-ray and TeV $\gamma$-rays that is located within the halo (left panel of Figure~\ref{fig:vela}). The halo is a diffuse nebula component observed brightly in radio and GeV bands. Due to the age of the system, the nebula components are likely the result of two particle populations: a high-energy, young (i.e., most recently injected by the pulsar) population that efficiently radiates in X-ray and TeV bands and a low-energy, old (i.e., earliest injected by the pulsar) population that efficiently radiates in radio and GeV bands. The short cooling times of the recently injected particles give rise to the compact size of the component, relative to the longer cooling times of the earlier injected particles, allowing for a more diffuse component. The different particle populations appear distinctly from the other in the broadband spectrum of the PWN, see the right panel of Figure~\ref{fig:vela}.

The distinct particle components arising in older PWNe is probably due to the evolutionary history, particularly the adiabatic and energetic losses suffered by the earliest injected particles, especially the radiative losses initiated during the passage of the reverse shock and which more recently injected particles may have not encountered by the time they enter the PWN. 

\subsection{PWN Evolution Effects on Radiative Properties}\label{sec:pwn_ev}

\begin{figure*}
\begin{minipage}[b]{1.0\textwidth}
\centering
\includegraphics[width=1.0\linewidth]{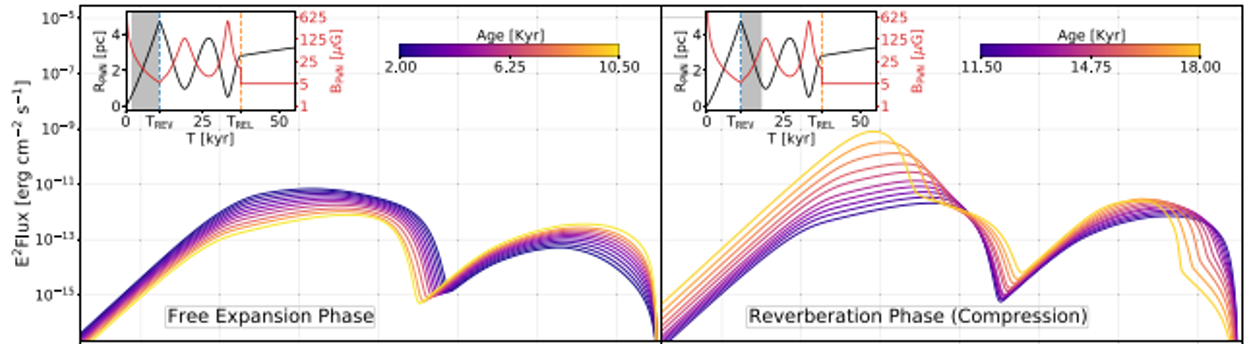}
\end{minipage}
\begin{minipage}[b]{.5\textwidth}
\centering
\includegraphics[width=1.0\linewidth]{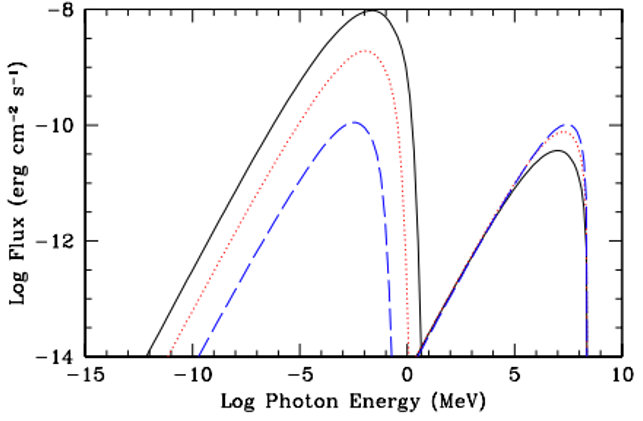}
\end{minipage}
\begin{minipage}[b]{.5\textwidth}
\centering
\includegraphics[width=1.0\linewidth]{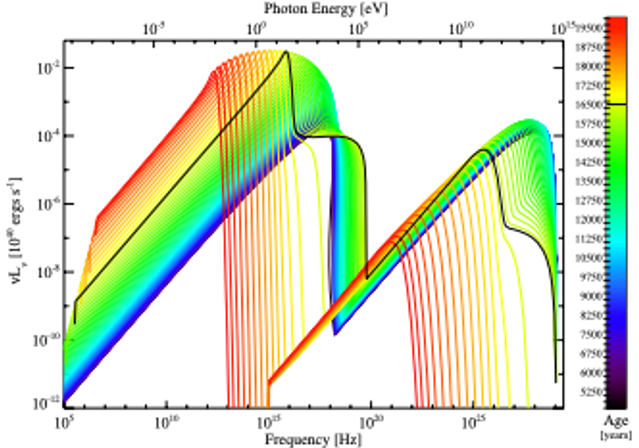}
\end{minipage}
\caption{{\it Left:} Simulated spectra of a PWN evolving in the free expansion phase. {\it Right:} Simulated spectra of a PWN evolving in the reverberation (or compression) phase. {\it Top Panels:} Adapted from \citep{fiori2022}. The inset in the top left corner shows the evolution of the PWN radius (black) and PWN magnetic field (red). {\it Bottom Left:} Adapted from \citep{slane2017}. The solid black curve corresponds to an age $\tau \sim 1$\,kyr, the red dotted line is for $\tau \sim 2$\,kyr, and the blue dashed line is for $\tau \sim 5$\,kyr. {\it Bottom Right:} Adapted from \citep{gelfand_2009}.}\label{fig:evolution_ex}
\vspace{-0.5cm}
\end{figure*}

Radiative models of PWN evolution inform our expectations for the detectability of PWNe across the broadband spectrum and can differ substantially depending on the age of the system \citep[see, for example,][]{gelfand_2009,torres2014,slane2017,fiori2022}. At early times, the PWN freely expands into the SNR interior. As a result of the increasing PWN volume, the magnetic field strength decreases with time during this phase while the ICS increases due to the continuous injection of relativistic particles from the central pulsar, see the left panels of Figure~\ref{fig:evolution_ex}. The free expansion phase ends when the SNR reverse shock reaches the PWN and depends on the ambient ISM density, but is generally expected to occur around $\tau \sim 5\,$kyr. When the SNR reverse shock begins interaction with the PWN, this marks the beginning of the reverberation (or compression) phase and can be a significant disruption to the PWN. The SNR reverse shock crushes the PWN, driving the pressure inside the PWN up as well as the magnetic field strength. The compression of the PWN by the reverse shock results in a radiative spectrum that reverses the trend from the free expansion phase: the synchrotron emission increases with the rapid rise in magnetic field strength while the ICS decreases due to the quick decline of the acceleration timescales, see the right panels of Figure~\ref{fig:evolution_ex}. 

\begin{figure*}
\begin{minipage}[b]{.5\textwidth}
\centering
\includegraphics[width=0.8\linewidth]{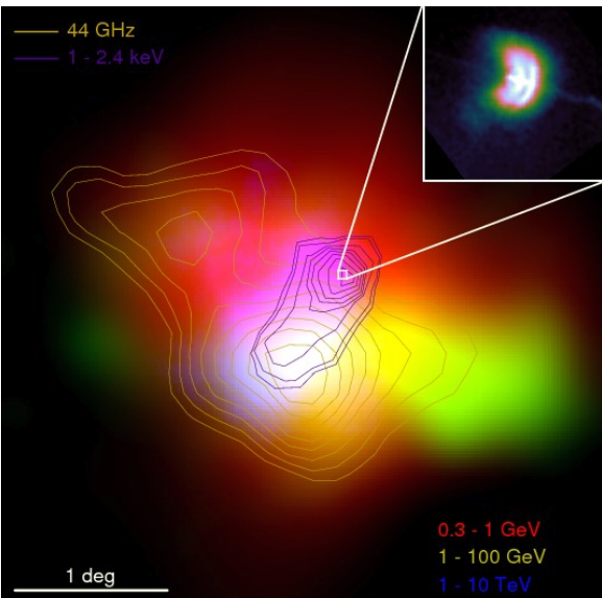}
\end{minipage}
\begin{minipage}[b]{.5\textwidth}
\hspace{-0.75cm}
\includegraphics[width=1.0\linewidth]{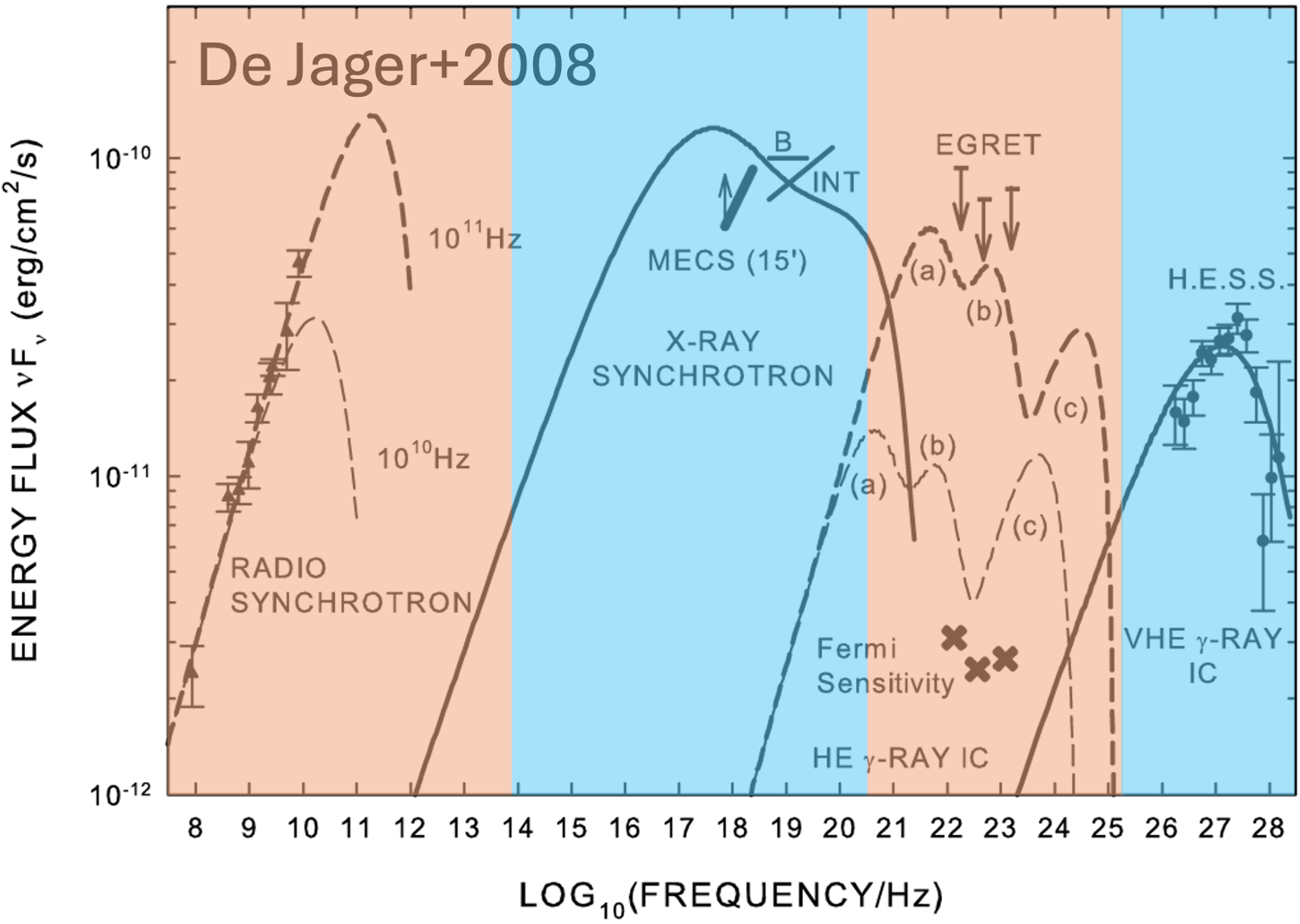}
\end{minipage}
\caption{{\it Left:} The broadband image of the Vela-X PWN, adapted from \citep{reynolds2017}. The 44\,GHz radio emission is shown as the yellow contours. 1--2.4\,keV X-ray emission is shown as the purple contours. The Chandra X-ray image of the innermost PWN and pulsar is shown as the upper right inset. The red, green, and blue colors of the main image represent the 0.3--1\,GeV, 1-100\,GeV, and 1--10\,TeV $\gamma$-ray emission. {\it Right:} The broadband data and model of the Vela-X PWN from \citep{dejager2008}. The low-energy population contribution is shaded in pink and the high-energy population contribution is shaded in blue. }\label{fig:vela}
\vspace{-0.5cm}
\end{figure*} 

The PWN compression phase is often also referred to as the reverberation phase, stemming from the asymmetry of PWNe and SNRs as they evolve in typically inhomogeneous ambient media that leads to a PWN-SNR reverse shock interaction that is also asymmetric. The PWN will re-expand and re-compress, or reverberate, several times during this phase until the pressure of the PWN and SNR interior achieve equilibrium. This phase does not end until the SNR enters its radiative phase, which can take several tens of kyrs. During the reverberation phase, it is possible for the pulsar to exit the nebula, enabled both by the velocity (magnitude and direction) of the pulsar and the direction of the initial passage of the reverse shock, determined by the ambient density profile. When the pulsar exits, the nebula becomes a relic of the oldest particles, radiating away any remaining energy and no longer receiving any energy input from the pulsar. The pulsar continues to inject high-energy particles into its surroundings, which generates a new, younger nebula that will concentrate in the immediate region of the pulsar and are radiatively limited by their short cooling times.

\begin{wrapfigure}{r}{0.5\textwidth}
\begin{minipage}{1.0\linewidth}
\vspace{-0.65cm}
\hspace{-0.35cm}
\includegraphics[width=1.0\linewidth]{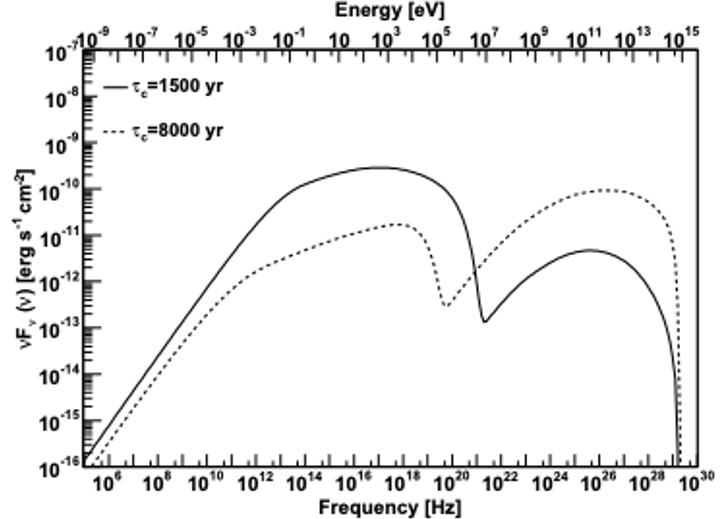}
\end{minipage}
\vspace{-.250cm}
\caption{A radiative model assuming Crab Nebula properties, changing only the value of the characteristic age of the system $\tau_c$, from 1500\,yr (solid line) to 8000\,yr (dashed line). Adapted from \citep{torres2014}.}\label{fig:torres_crab}
\vspace{-0.5cm}
\end{wrapfigure}

Since PWN evolution is strongly influenced by the central pulsar, host SNR, and the surrounding medium, the radiative properties heavily depend on the system age, magnetic field strength of the PWN, the pulsar spin down power, and the structure of the ambient medium. The age of the system alone, already evidenced by the models described above, can be a deciding parameter in the observed spectrum, as demonstrated in the Crab Nebula radiative model presented in \citep{torres2014} and shown in Figure~\ref{fig:torres_crab}.

In summary, the spectral signatures of a PWN can vary widely for various reasons. Young and old PWNe can exhibit complex particle properties involving multiple populations and consequent emission regions. As shown for the Crab Nebula ($\tau \sim 1\,$kyr) in Figure~\ref{fig:crab_spec} compared to the Vela-X nebula ($\tau \sim 10\,$kyr) in Figure~\ref{fig:vela}, right panel, it is possible to have multiple emission components generated by different acceleration mechanisms occurring in different regions of the PWN (e.g., the Crab) and from multiple particle populations manifesting from the compression of the PWN by the SNR reverse shock (e.g., Vela-X). As a consequence, it is important to consider the properties of the entire broadband spectrum of a PWN in addition to the physical properties of the pulsar-PWN-SNR system to adequately characterize the underlying particle populations. 

\subsection{$\gamma$-ray Signatures}\label{sec:gamma-ray_sigs}

\begin{figure*}
\begin{minipage}[b]{1.0\textwidth}
\centering
\includegraphics[width=0.7\linewidth]{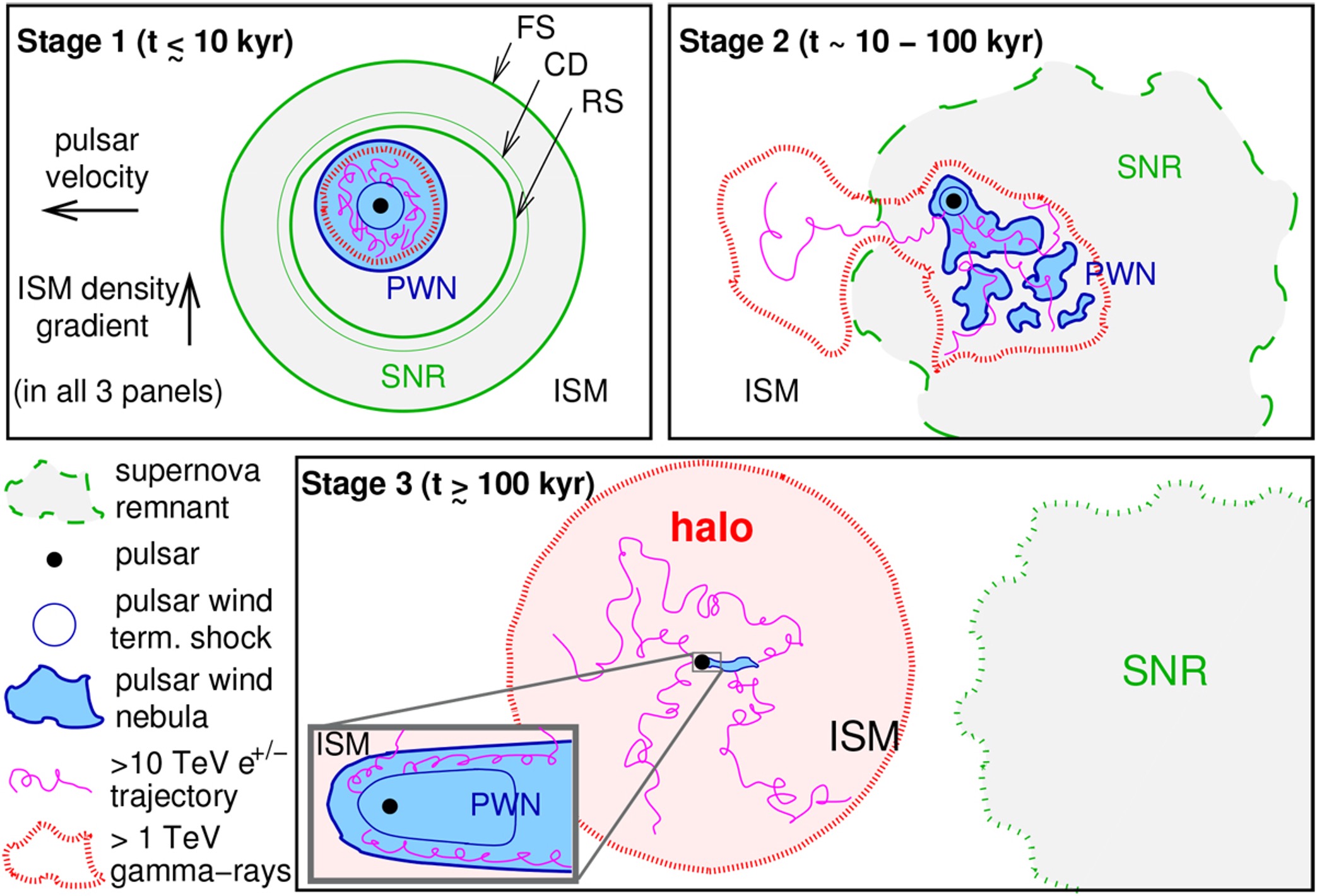}
\end{minipage}
\caption{A graphic illustrating the stages of PWN evolution, adapted from \citep{giacinti2020}. Stage 1 represents the free expansion phase. FS = forward shock, CD = contact discontinuity, and RS = reverse shock. Stage 2 represents the later times of the reverberation phase where the highest-energy particles begin to diffuse into the SNR interior and ISM. Stage 3 represents the latest evolutionary stage of the PWN, where it becomes a halo of diffuse high-energy particles interacting with the ISM entirely, having exited the SNR. As the pulsar moves through the ISM, it may become supersonic, which generates a compact bow-shock nebula in the immediate region of the pulsar (the bottom left inset).}\label{fig:giacinti}
\vspace{-0.5cm}
\end{figure*}

The final stages of PWN evolution ($ 5 \leq \tau \leq 100$\,kyr), well after the onset of the reverberation phase, will see the weakest magnetic field strength but the largest particle population, leading to weak synchrotron emission but bright ICS emission. Therefore, ICS in the $\gamma$-ray regime is expected to be the most efficient way to detect PWNe, see Figure~\ref{fig:giacinti} and \citep{giacinti2020}. The highest-energy particles of an evolved PWNe will begin to escape the confinement of the PWN and begin interacting with the SNR interior and ISM. If the pulsar exits the SNR entirely, it may become supersonic as it moves through the ISM, which generates a bow-shock nebula in its immediate surroundings. The highest-energy particles continue diffusing into the ISM, generating a diffuse halo of ICS emitting particles. Called TeV halos, they represent some of the oldest $\tau \gtrsim 100\,$kyr PWNe and are markedly different from other stages of a PWN since the diffused PWN particles now interact with the ISM. 

\begin{figure*}
\centering
\includegraphics[width=0.32\linewidth]{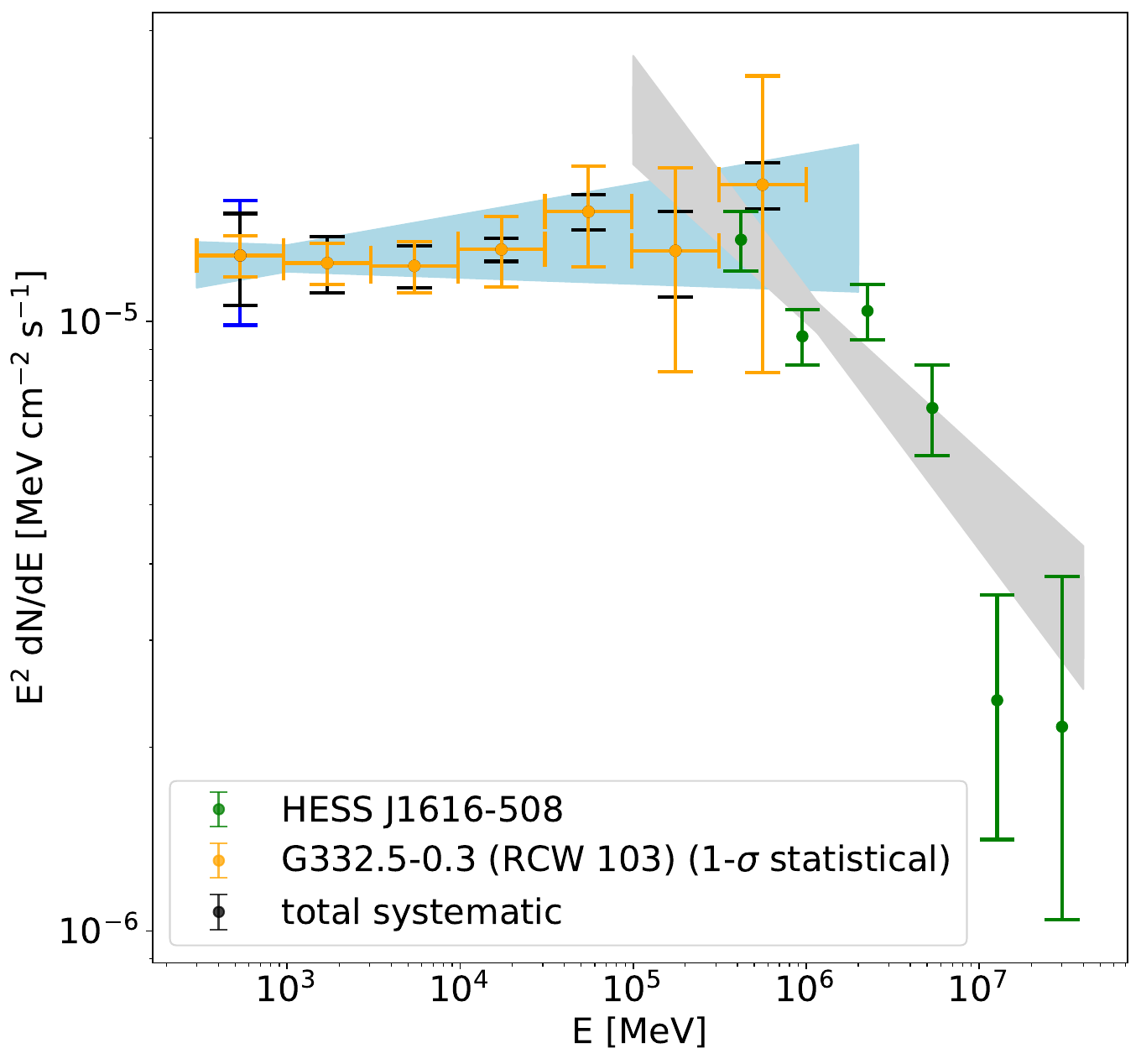}
\includegraphics[width=0.32\linewidth]{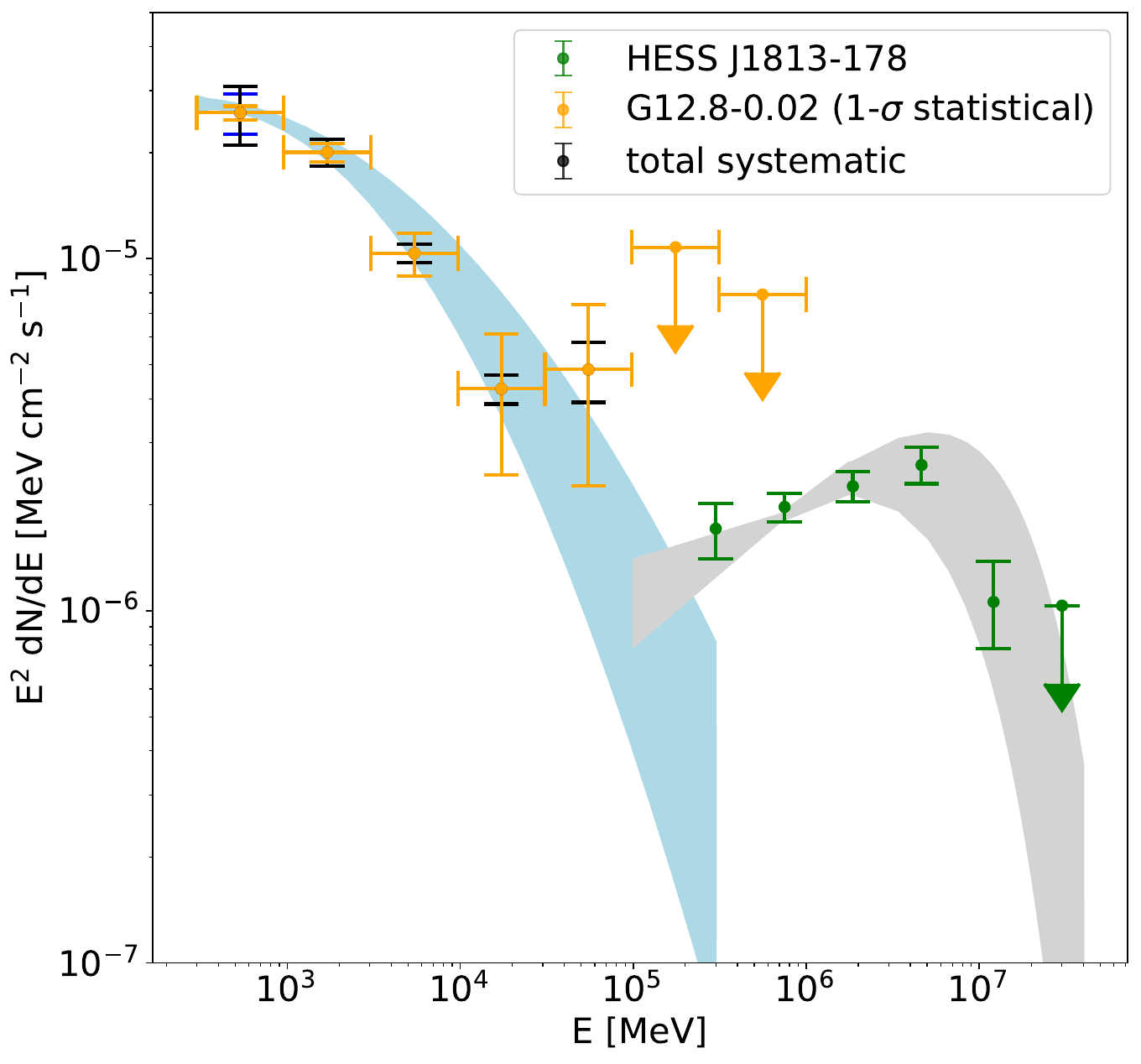}
\includegraphics[width=0.32\linewidth]{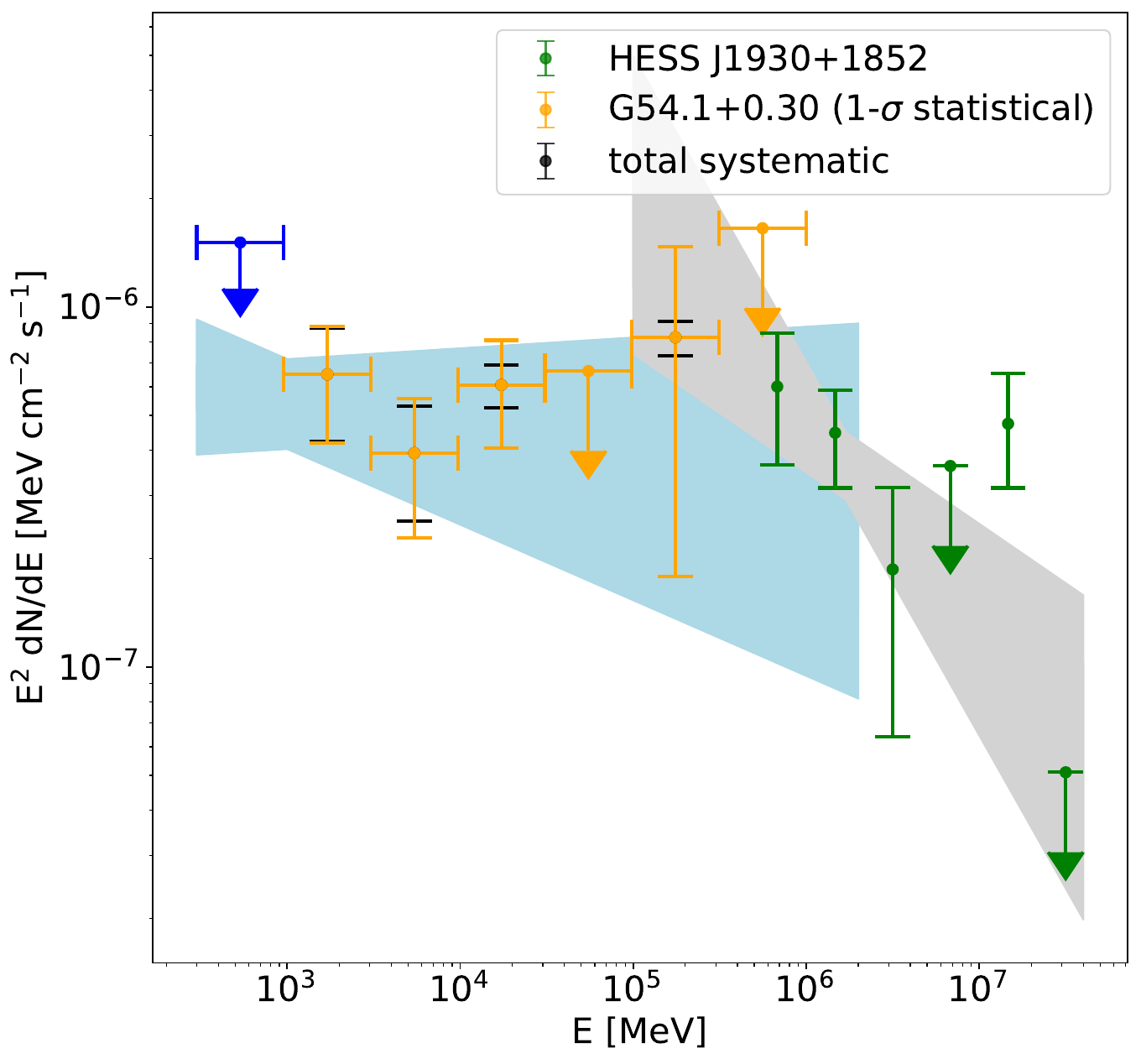}
\includegraphics[width=0.32\linewidth]{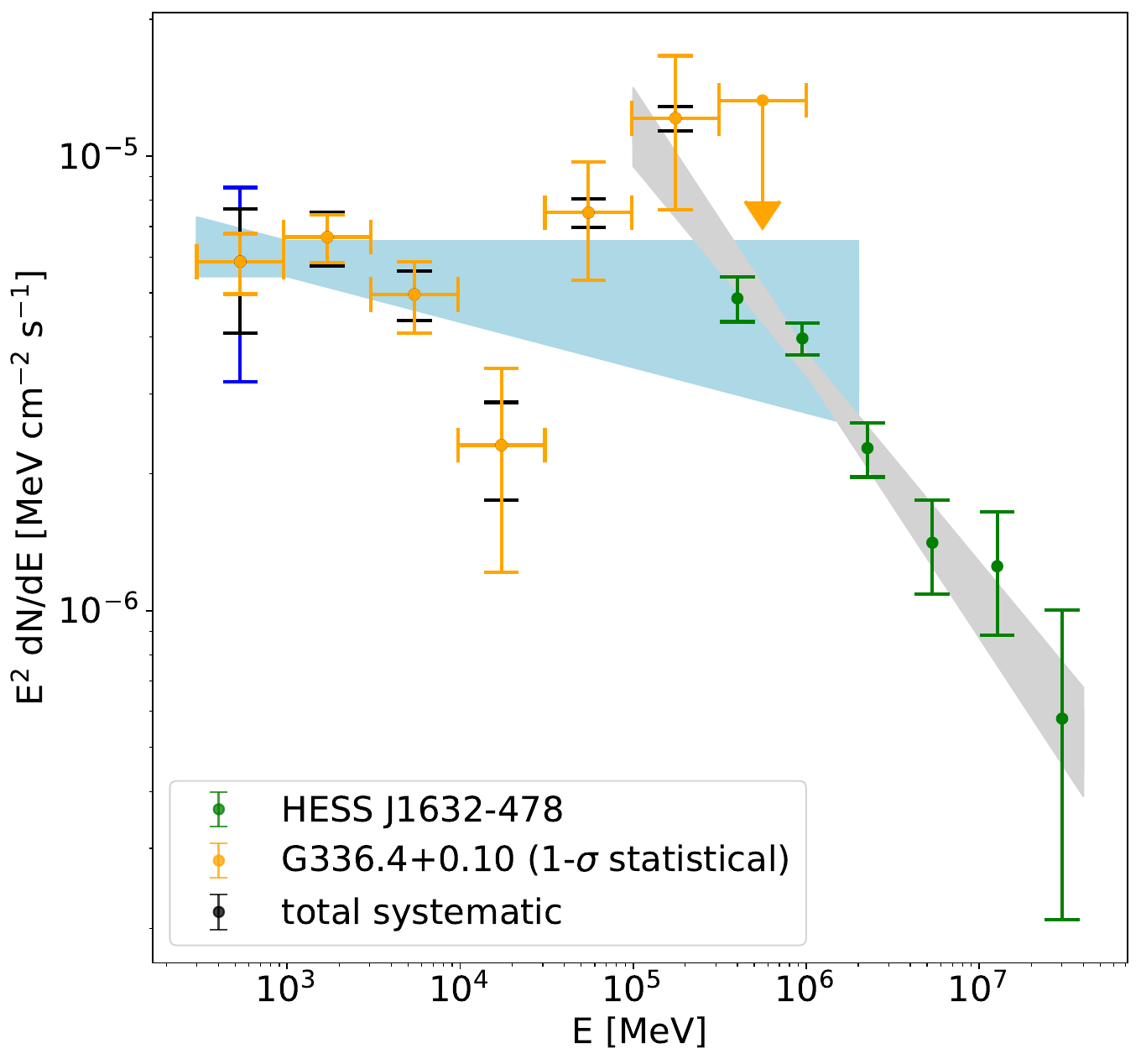}
\includegraphics[width=0.32\linewidth]{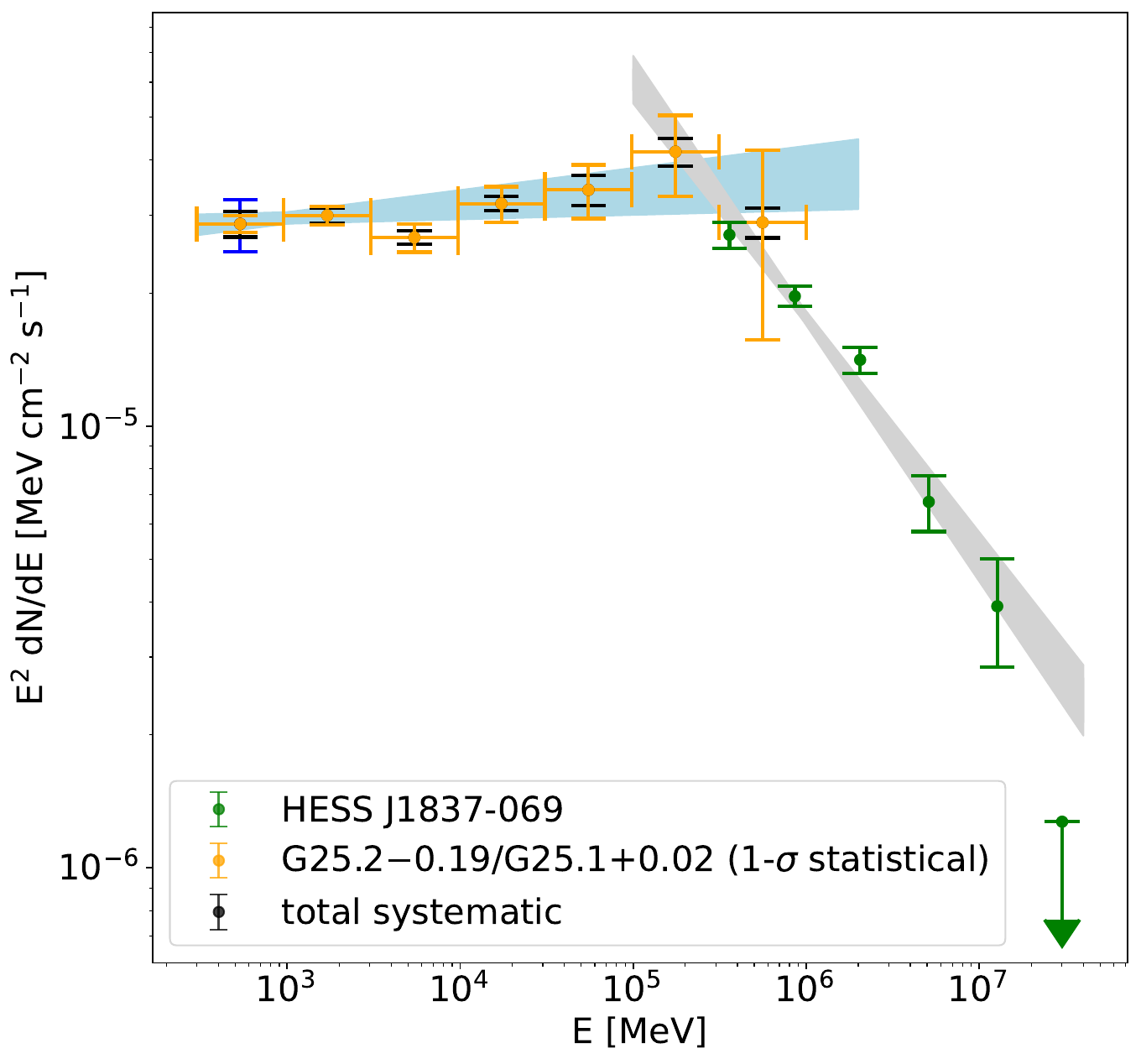}
\caption{A sample of source spectra from Fermi--LAT data (yellow points with systematic errors in black and blue) and HESS data (green points) from \citep{eagle2025}. Spectral models are shown as the shaded bands for Fermi--LAT (blue) and HESS (grey). The age estimates for the sample range from $3\,$kyr (G54.1+0.3) to $22.7\,$kyr (HESS~J1837--069), based on characteristic ages compiled in \citep{eagle2025}.}
\label{fig:eagle_models}
\end{figure*}

Due to their high likelihood of detection, evolved PWNe and TeV halos have become prominent targets for $\gamma$-ray searches in both the GeV and TeV bands \citep[e.g.,][]{acero2013, eagle2025}. Further, the majority of Galactic TeV sources have been identified as PWNe \citep[e.g.,][]{hessgps2018}. Several TeV PWNe indicate that their ICS $\gamma$-ray peaks are occurring $E < 1\,$TeV, which is where the Fermi--Large Area Telescope (LAT) is the most sensitive, see Figure~\ref{fig:eagle_models}, for a sample of TeV sources and their $\gamma$-ray spectra, taken from \citep{eagle2025}.

\section{$\gamma$-ray Searches}\label{sec:gamma_searches}

\begin{wrapfigure}{l}{0.6\textwidth}
\begin{minipage}{1.0\linewidth}
\centering
\includegraphics[width=1.0\linewidth]{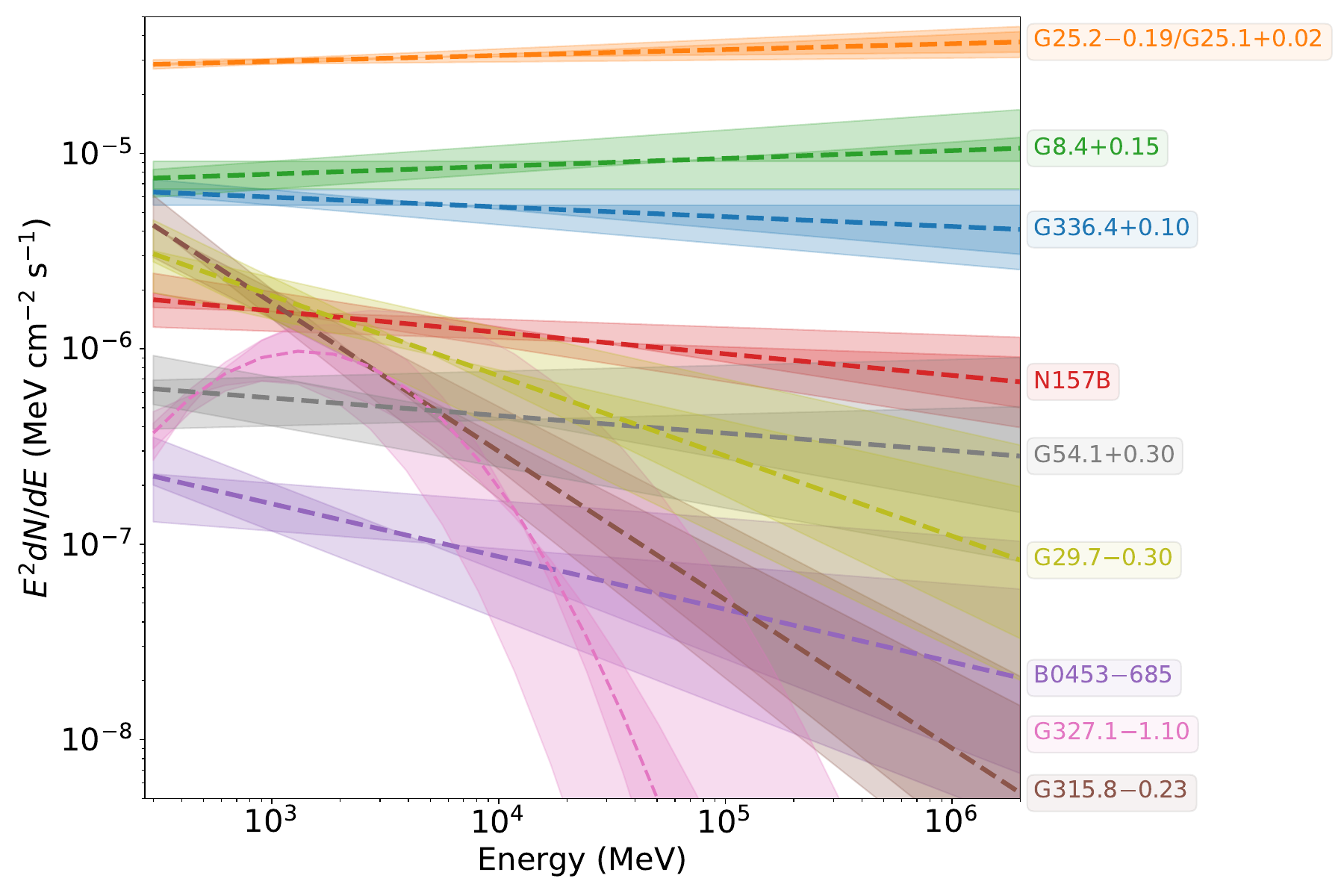}
\end{minipage}
\caption{The best-fit spectral models of the 9 likely LAT PWNe identified in \citep{eagle2025}.}\label{fig:eagle_2025}
%\vspace{-0.5cm}
\end{wrapfigure}

The Fermi--LAT has discovered a total of 12 PWNe and another 9 as PWN associations since being launched in 2008 \citep{4fgl-dr4}. Several of the 12 LAT PWNe were discovered from analyzing the off-pulse emission of Fermi--LAT detected pulsars \citep[e.g., the Crab, Vela-X and 3C 58,][]{ackermann2011,2fgl,2pc}. Others were identified for energies $E> 10\,$GeV, targeting the locations of TeV sources \citep{acero2013}. Most recently, \citep{eagle2025} performed a GeV search with the Fermi--LAT targeting the locations of all known PWNe that have been identified across the broadband spectrum and lack a Fermi--LAT detected pulsar. The GeV search of \citep{eagle2025} finds 9 likely LAT PWNe and another 21 as PWN candidates. The best-fit spectral models of the 9 likely PWNe are shown in Figure~\ref{fig:eagle_2025}.

%\begin{wrapfigure}{r}{0.6\textwidth}
%\begin{minipage}{1.0\linewidth}
%\centering
%\includegraphics[width=1.0\linewidth]{acero_2013_tev_peak.png}
%\end{minipage}
%\caption{A sample of TeV sources (blue data) and their broadband spectra including the Fermi--LAT measurements (black data) from \citep{acero2013}.}\label{fig:acero_2013}
%\vspace{-0.5cm}
%\end{wrapfigure} 
%replace with my own:

%RCW 103 tau_c = 8.3kyr
%J1813-178 tau_c = 5.6kyr
%G54.1+0.3 tau_c = 2.89kyr
%J1632-478 tau_ ~ 10kyr from Balbo+2010
%J1837-069 tau_c = 22.7kyr

\begin{figure*}[b]
\begin{minipage}[b]{0.5\textwidth}
\centering
\includegraphics[width=0.9\linewidth]{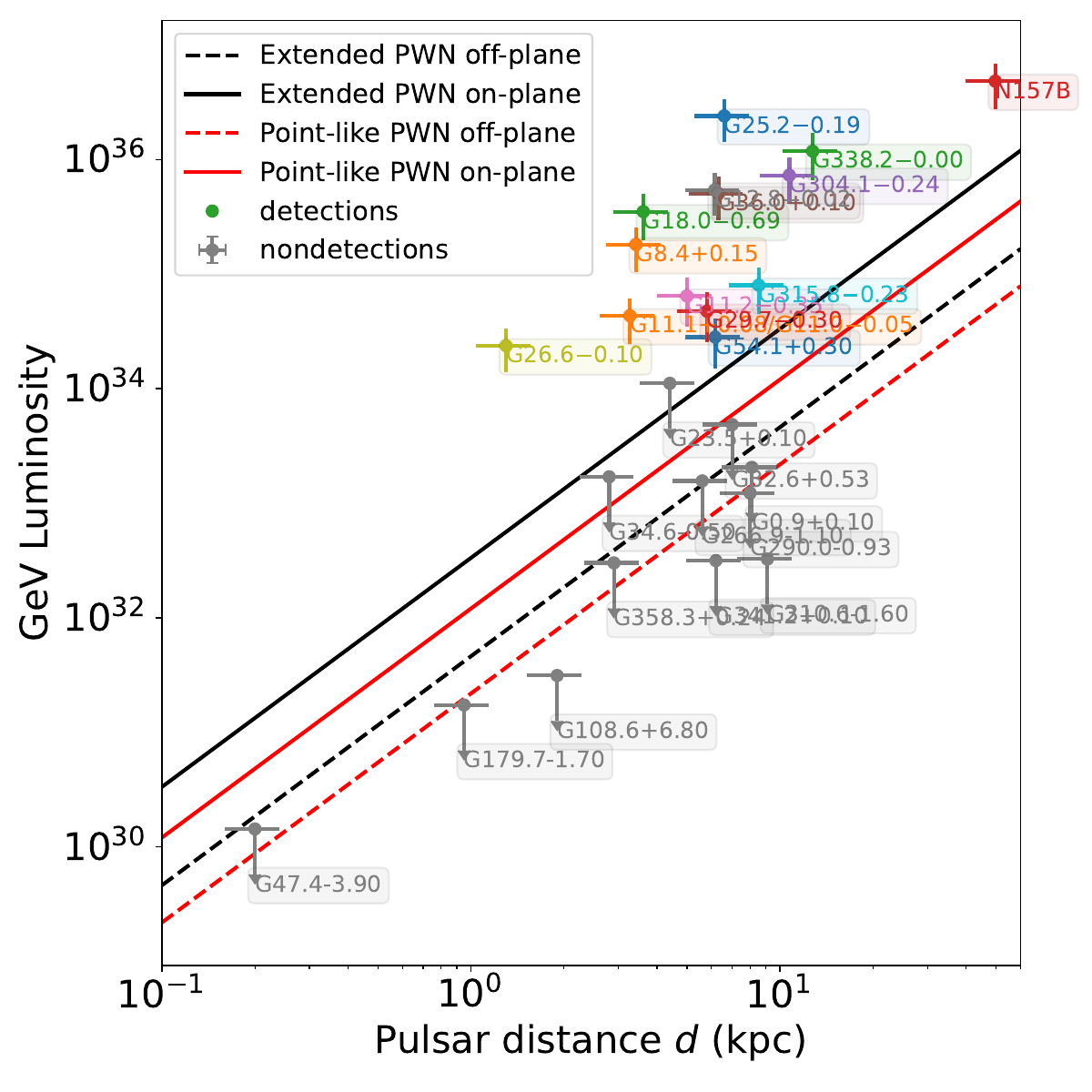}
\end{minipage}
\begin{minipage}[b]{0.5\textwidth}
\centering
\includegraphics[width=0.95\linewidth]{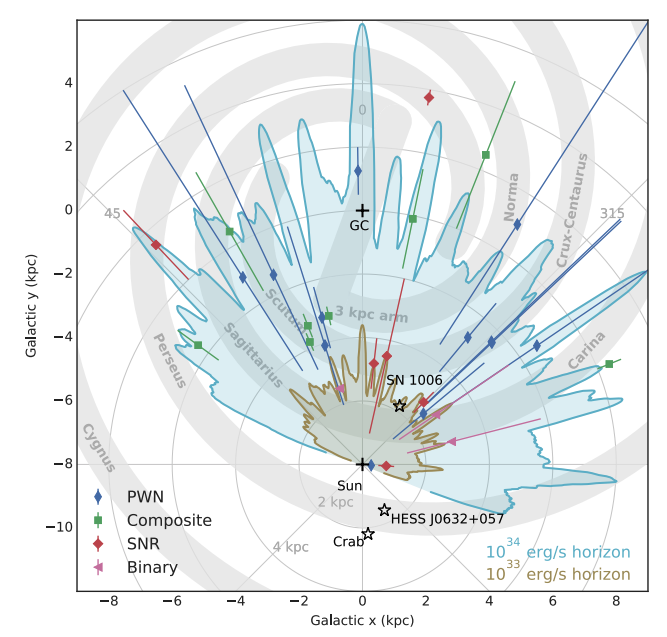}
\end{minipage}
\caption{{\it Left}: The GeV luminosity as a function of pulsar distance for detected LAT sources (colored points) and undetected sources (grey upper limits) compared to the 12-yr Fermi--LAT flux sensitivity curves from \citep{eagle2025}. {\it Right:} The TeV luminosity horizon of the HESS Galactic Plane Survey, demonstrating the constraints to detecting PWNe within a certain distance that have the luminosity of the corresponding shaded areas, from \citep{hessgps2018}.}\label{fig:gamma_sigs}
\vspace{-0.5cm}
\end{figure*}

The results of the recent GeV search \citep{eagle2025} and TeV search \citep{hessgps2018} are limited to the brightest, closest PWNe, illustrated by the GeV luminosity as a function of pulsar distance in the left panel of Figure~\ref{fig:gamma_sigs} and by the luminosity horizon (shaded regions) in the right panel of Figure~\ref{fig:gamma_sigs}. 
%consider removing this paragraph and its corresponding figure?
While the TeV $\gamma$-ray band has led to an increasing discovery of PWNe and their spectral characteristics (36 listed as PWNe in the TeV Catalog \citep{tevcat2008}), compared to their radio and X-ray counterparts ($\gtrsim 30$ in radio and $\gtrsim 60$ in X-ray), the accurate identification and characterization of recent GeV PWN associations \citep{eagle2025} as well as the new $E>1\,$TeV and $E>100\,$TeV PWN associations \citep{lhaaso2024} rely on the next generation of $\gamma$-ray observatories. 

\begin{figure*}
\begin{minipage}[b]{0.5\textwidth}
\centering
\includegraphics[width=1.\linewidth]{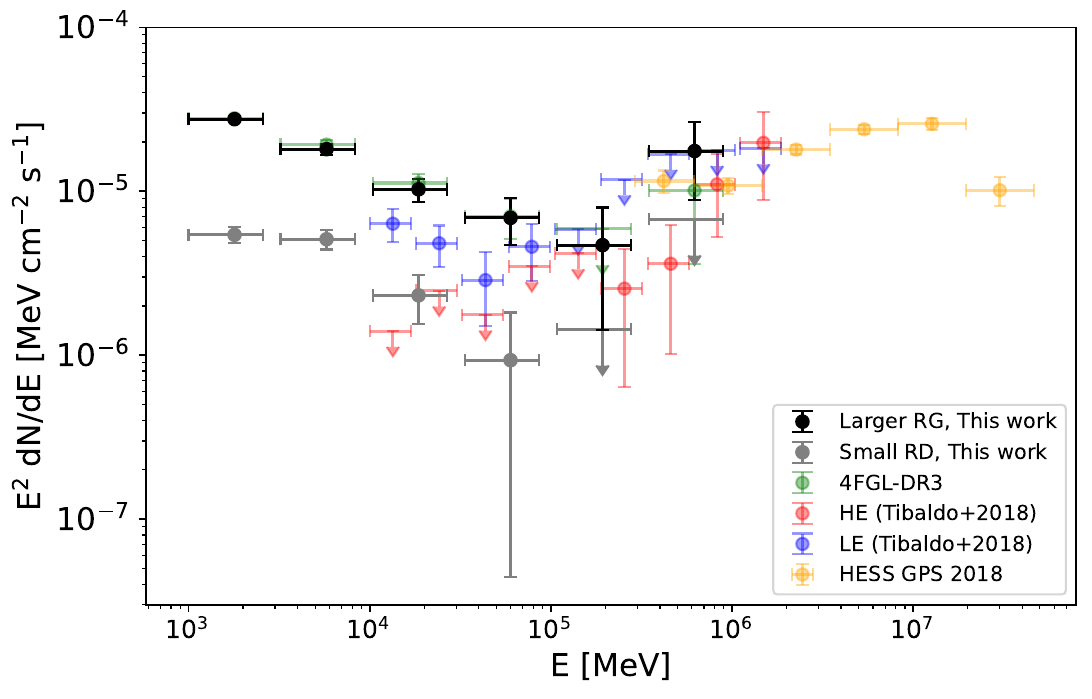}
\end{minipage}
\begin{minipage}[b]{0.5\textwidth}
\centering
\includegraphics[width=0.9\linewidth]{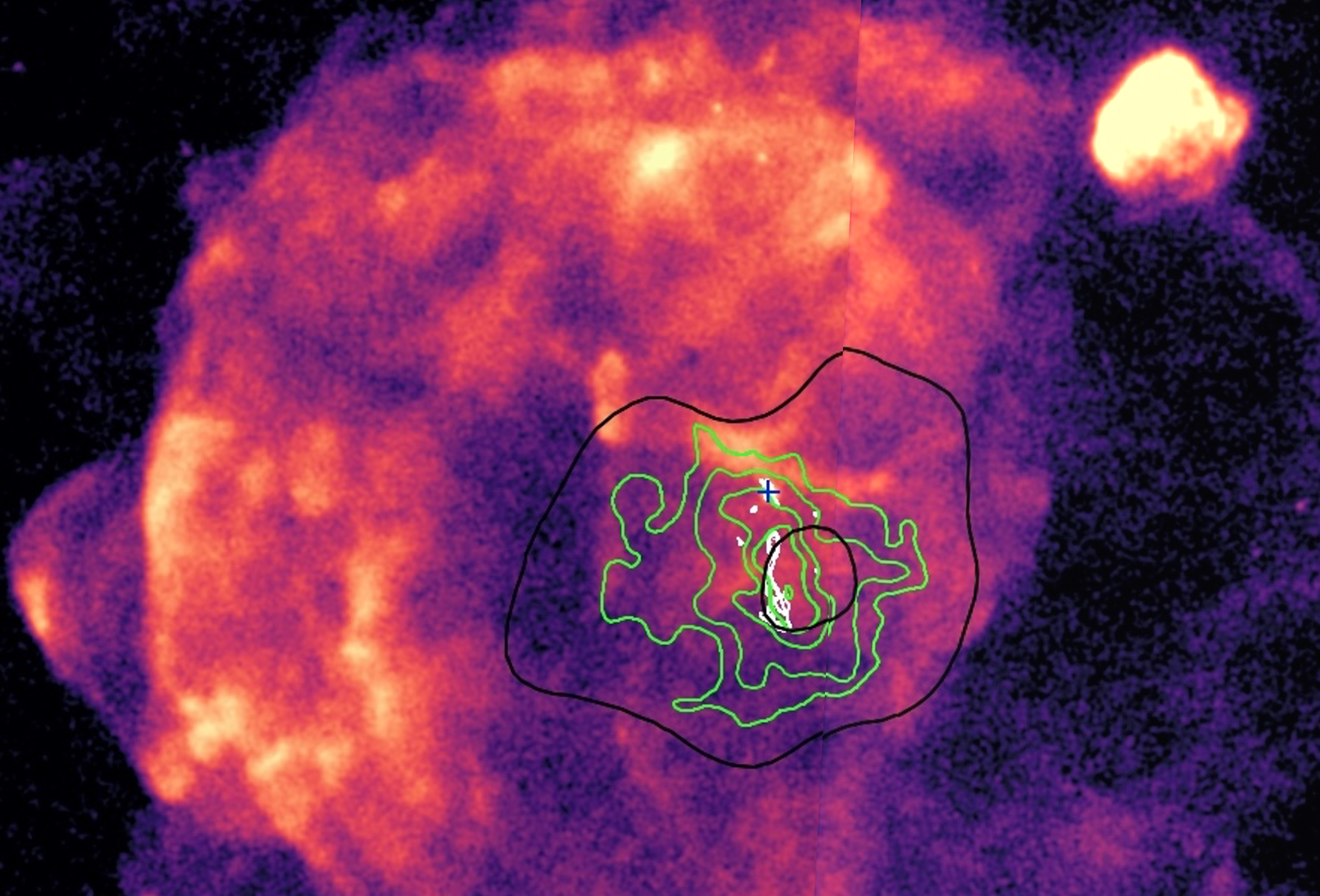}
\end{minipage}
\caption{{\it Left:} The $\gamma$-ray spectral data of the Vela-X PWN reported in \citep{lange2025}. {\it Right:} The 1990 ROSAT sky map of the Vela SNR and PWN in the soft X-ray band ($E<2.4\,$keV). The 330\,MHz radio PWN is outlined by the outer and inner black contour representing the minimum and maximum radio flux values, respectively. The TeV PWN flux is shown as the green contours. The white contours represent the 843\,MHz radio flux of the PWN. A blue cross marks the position of the pulsar.}\label{fig:pwn_examples}
\vspace{-0.5cm}
\end{figure*}

As an example, the $\gamma$-ray spectrum of Vela-X is shown in the left panel of Figure~\ref{fig:pwn_examples}, which shows a clear increase in flux towards MeV energies, similar to the MeV synchrotron bump in the Crab spectrum observed by COMPTEL (Figure~\ref{fig:crab_spec}). The shape of the $\gamma$-ray spectrum highlights the importance of understanding the underlying particles and emission components, as there is a clear indication for multiple particle populations and/or multiple emission components contributing to the GeV--TeV spectrum. A similar complexity is observed in the spatial morphology of the Vela-X PWN, shown in the right panel of Figure~\ref{fig:pwn_examples}. A diffuse nebula component is detected in 330\,MHz radio and Fermi--LAT GeV bands that spans several degrees in size, while more compact 843\,MHz radio, X-ray, and TeV emission concentrates closer to the pulsar. The spatial morphology is compelling evidence that the reverse shock has compressed the PWN, leading to multiple particle populations undergoing different energy losses. 

The energy-dependent morphology and spectral features is observed in many PWNe, which have also been detected in the $\gamma$-ray band. B0453--685, an evolved $\tau \sim 14$\,kyr composite SNR located in the Large Magellanic Cloud (LMC), was recently detected by the Fermi--LAT \citep{eagle2023}. The radio to $\gamma$-ray radiative models require two electron populations to explain the broadband emission from the PWN, and additionally predicts an MeV bright pulsar contributing $E<5\,$GeV to the Fermi--LAT spectrum. The central pulsar of B0453--685 has remained undetected in any waveband, but if it is MeV bright, it may be easily detectable by a future MeV mission. Similar observations for the PWN G327.1--1.1 ($\tau \sim 18$\,kyr) depict an evolved PWN disrupted by the SNR reverse shock, with a unique displacement between the radio and X-ray nebula components, and provides an excellent example of a relic, radiating primarily in the radio, and a young nebula, radiating primarily in the X-ray, generated by the exiting pulsar \citep{temim2015}. The different particle components are further evident in the PWN broadband spectrum of G327.1--1.1 where the recent Fermi--LAT $\gamma$-ray detection hints at a potential pulsar component below $E \lesssim 10\,$GeV \citep{eagle2022}. Finally, the case of MGRO~J1908+06 also points to an evolved PWNe, powered by an energetic pulsar that has a characteristic age $\tau_c \sim 20\,$kyr, and exhibits energy-dependent morphology in the $\gamma$-ray band as observed by the Fermi--LAT, VERITAS, and HAWC observatores \citep{j1908_2024}. This is not an exhaustive list of $\gamma$-ray PWNe presenting interesting broadband observational spectral or spatial features \citep[see e.g., HESS~J1303--631, HESS~J1825--137, MSH~15--56,][for other examples]{hess2012,principe2020,devin2018}. 

\section{Next Generation of $\gamma$-ray Observatories}\label{sec:gamma_future}

The next generation of $\gamma$-ray observatories will expand the discovery space for PWN detection and spectral characterization. Future TeV observatories such as the Cherenkov Telescope Array Observatory (CTAO) \citep{cta2023} will be able to detect many more PWNe up to flux limits $\sim10^{-15}$ TeV cm$^{-2}$ s$^{-1}$, increasing the detection sensitivity by $\sim 3$ relative to the HESS Galactic Plane Survey \citep{hessgps2018}, which can detect PWNe up to a distance of 7.3\,kpc for a TeV luminosity $L_\gamma \sim 10^{34}$ erg s$^{-1}$ (Figure~\ref{fig:gamma_sigs}, right panel). The CTA will also achieve a better angular resolution than MAGIC, VERITAS, and HAWC observatories, and especially that of the LAT, reaching a containment angle $\sim 0.05\,\degree$ at 1\,TeV that narrows slightly at increasing energies. The Southern Wide-field Gamma-ray Observatory (SWGO) \citep{petra2019} will also provide competitive flux sensitivity to the CTA, reaching a limit $\sim 10^{-13}$\,erg cm$^{-2}$ s$^{-1}$ at $E = 100\,$TeV, and providing important overlap between HAWC and LHAASO flux coverage. The improvements in the TeV band will not only increase the detected PWN population, but will be able to provide spatial constraints not currently possible, such as resolving PWN TeV morphology to determine more precisely the association between radio and X-ray components, enabling a better understanding of the nature of associated GeV emission \citep[e.g.,][]{eagle2022}.

\begin{figure*}
\begin{minipage}[b]{0.35\textwidth}
\hspace{-0.65cm}
\includegraphics[width=1.15\linewidth]{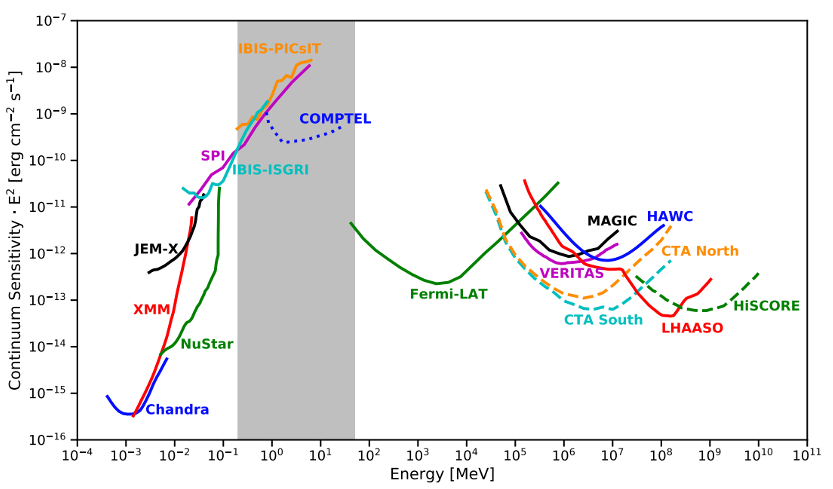}
\end{minipage}
\begin{minipage}[b]{0.25\textwidth}
\centering
\includegraphics[width=1.0\linewidth]{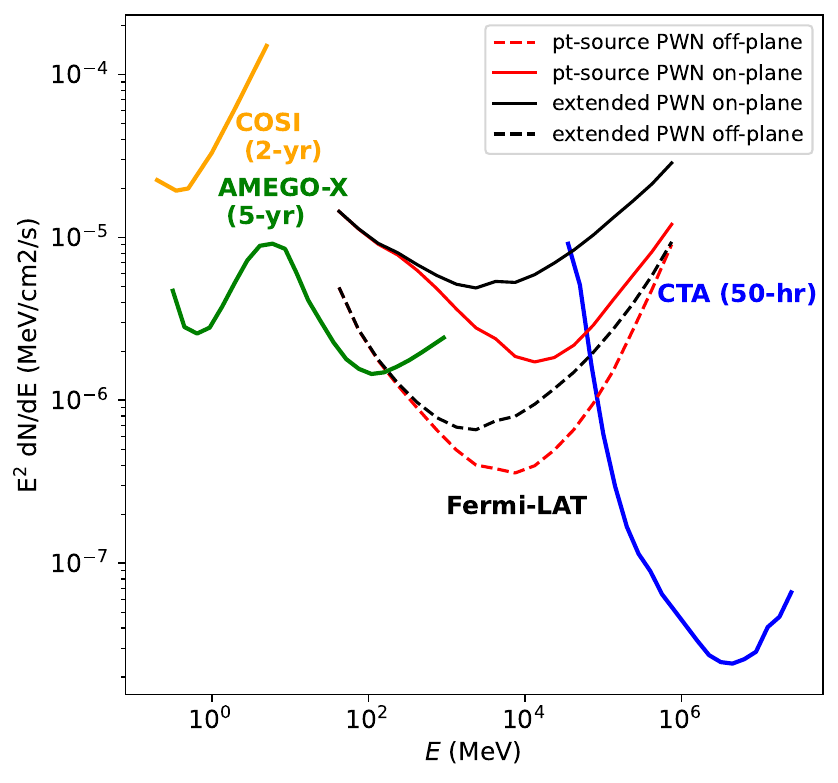}
\end{minipage}
\begin{minipage}[b]{0.4\textwidth}
\centering
\includegraphics[width=1.15\linewidth]{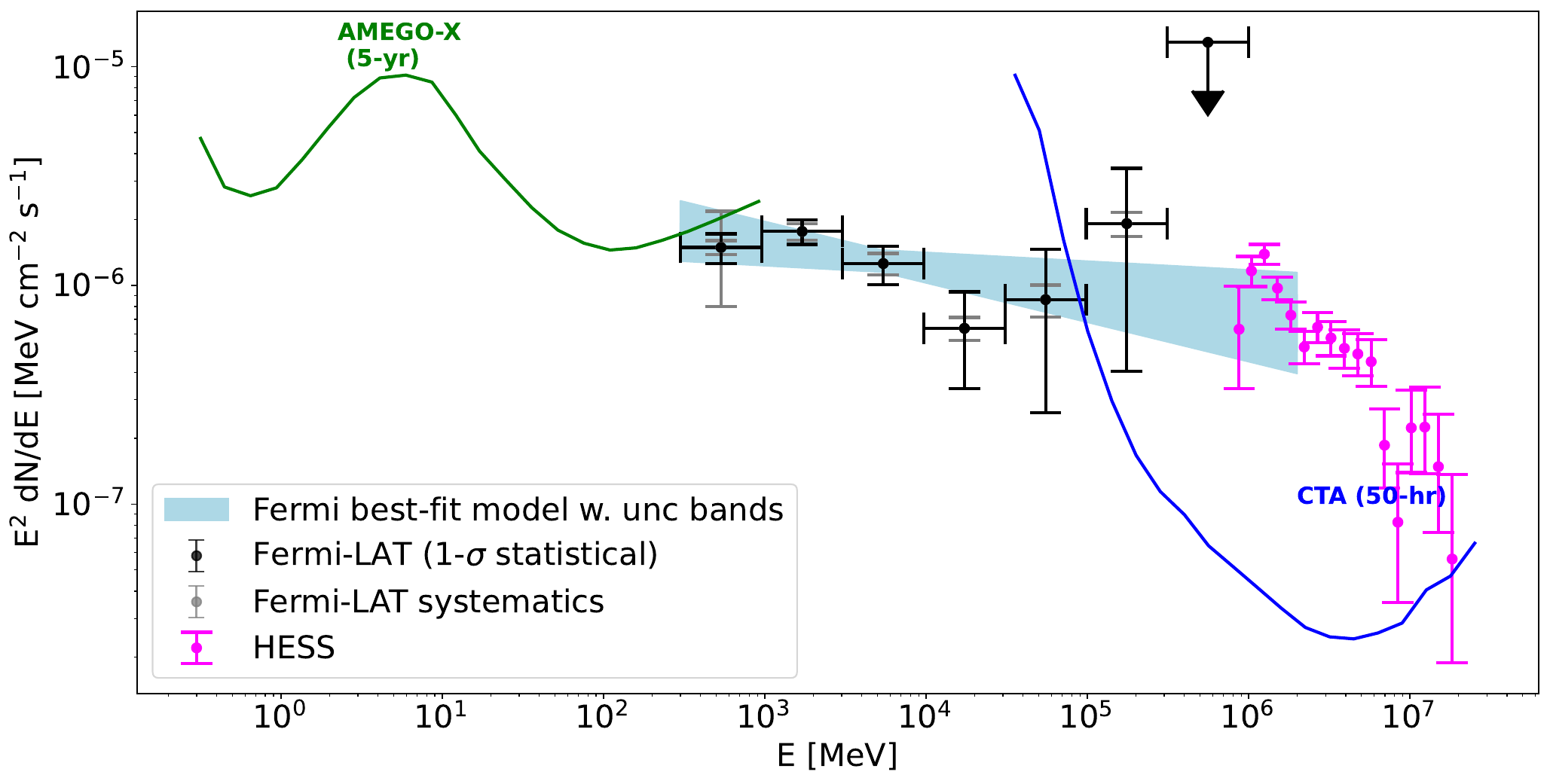}
\end{minipage}
\caption{{\it Left:} Flux sensitivity curves of various instruments from \citep{lucchetta2022}, highlighting in grey the least explored bandpass, the MeV band. {\it Middle:} The 12-yr Fermi--LAT flux sensitivity curves compared to the future MeV missions COSI and AMEGO-X and the future TeV observatory CTA. {\it Right:} The Fermi--LAT data and model of the PWN N157B from \citep{eagle2025} along with the TeV data from \citep{hess_n157b} and AMEGO-X and CTA flux sensitivity curves.}\label{fig:future}
\vspace{-0.5cm}
\end{figure*} 

Furthermore, future instruments that will explore the MeV bandpass (see Figure~\ref{fig:future}) will add crucial information to these systems including the detection of MeV bright pulsars and PWNe, and open up the possibility for new discoveries of MeV flaring in PWNe. The Crab Nebula is the only PWN with detected MeV and GeV flares, thought to originate from magnetic reconnection occurring within the nebula \citep{cerutti2012,arakawa2020}. A similar energetic PWN located in the LMC, N157B, is a promising candidate of MeV flaring \citep[e.g.,][]{saito2017}. The acceleration mechanisms present in a PWN determine the maximum particle energy and how it evolves with time which in turn has great impact on the MeV luminosity evolution of a PWN, due to the short synchrotron lifetime of the highest energy particles \citep[e.g.,][]{gelfand2019}. Therefore, accurate spectral characterization of the ICS emission from PWNe will be enabled by future MeV observations, adding valuable constraints to the particle properties and their emission, see the middle and right panels of Figure~\ref{fig:future}, see also \citep{eagle_frascati,eagle2024}.

Future observatories in the MeV and TeV bands, together with the Fermi--LAT and current TeV telescopes, will significantly enhance the understanding of energetic environments in the Galaxy such as PWNe, providing an unprecedented view of efficient particle acceleration processes, sites, and conditions, which remain beyond reach without.

\section{Conclusion}\label{sec:conclusion}
We have described the current observational knowledge and theoretical basis for PWNe that identify these sources as excellent targets to understand particle acceleration in extreme conditions within the Milky Way Galaxy and the capacity for sources like PWNe to explain the leptonic contribution to the Galactic CR flux. The radiative properties of PWNe from radio to TeV $\gamma$-ray bands provide valuable insight into the acceleration mechanisms and regions present within them and emphasize that the evolutionary history of these systems is important to accurately determine the underlying particle properties. PWNe are now well known to be efficient $\gamma$-ray emitters, giving way to high-energy studies as the key to understanding the relativistic particles and their environments. We have outlined the physical traits of a PWN that can produce particle acceleration and explain the broadband spectral and spatial observations. Finally, we underscore the importance of future observatories in the $\gamma$-ray band that will provide the necessary constraints to the spectral characterization and the underlying particle properties of PWNe. Continued progress in $\gamma$-ray astrophysics reflects the significant efforts of the community, ensuring a $\gamma$-ray bright future.

\bigskip
\bigskip
\noindent {\bf DISCUSSION}

\bigskip
\noindent {\bf BIRENDRA CHHOTARAY:} Are PWNe found in X-ray binaries? If yes, how does the PWN affect the companion?

\bigskip
\noindent {\bf JORDAN EAGLE:} Here, we do not discuss PWNe that are part of binary systems, but in general, yes, there are PWNe in binary systems including X-ray binaries. In these systems, the companion is largely the one that affects the PWN. The photon field of the stellar companion as well as its magnetic field serve as the main environment for producing synchrotron and ICS radiation in the PWN. 

\end{document}